\documentclass[11pt]{article} 
\usepackage{amsmath,amsthm,amsfonts,amssymb,mathrsfs,epsfig,bm}
\usepackage{pxfonts}
\usepackage{microtype}
\usepackage[usenames,dvipsnames]{xcolor}
\usepackage[colorlinks=true]{hyperref}
\usepackage{graphicx}
\usepackage[margin=1cm]{caption}

\usepackage{caption}
\usepackage[top=2cm, bottom=2cm, left=2cm, right=2cm, heightrounded,
marginparwidth=2.9cm, marginparsep=2mm]{geometry}
\parskip        \smallskipamount

\newtheorem{remark}{Remark}

\def\Z{\mathbb{Z}}

\def\R{\mathbb{R}}

\def\P{\mathbb{P}}
\def\E{\mathbb{E}}
\renewcommand{\phi}{\varphi}
\renewcommand{\epsilon}{\varepsilon}
\def\BB{\mathcal{B}}
\def\PP{\mathcal{P}}
\newcommand{\1}{\boldsymbol{1}}

\newcommand{\eqdist}{\stackrel{\text{\rm (d)}}{=}}

\def\no{\noindent}

\def\hG{\hat{G}}
\usepackage{comment}
\def\sP{\mathsf P}

\usepackage{mathtools}

\begin{document}


\title{\bf Age of information for small buffer systems}
\author{George Kesidis\thanks{gik2@psu.edu; Pennsylvania State Univ., USA}
\and Takis Konstantopoulos\thanks{takiskonst@gmail.com; Univ.\ of Liverpool, UK; research supported by Cast.\ Co.\  IIS-75}
\and Michael A.\ Zazanis\thanks{zazanis@aueb.gr; Athens Univ.\ of Economics and 
Business, Greece}
}
\date{\small 14 June 2021}
\maketitle

\begin{abstract}
Consider a message processing system whose objective is to produce
the most current information as measured by the quantity known
as ``age of information''. We have argued in previous papers
that if we are allowed to design the message processing policy {\em ad libitum}, 
we should keep a small buffer and operate according to a LIFO policy.
In this small note we provide an analysis for the AoI of the $\PP_m$ system
which uses a buffer of size $m$, a single server, operating without service
preemption and in a LIFO manner for stored messages.
Analytical expressions for the mean (or even distribution) of the AoI in steady-state
are possible but with the aid computer algebra. We explain the the analysis for
$m=3$.
\end{abstract}


\section{The background}
Here is the definition of the Age-of-Information (AoI) stochastic process $\alpha(t)$,
$t \in \R$, for a fairly general message processing system.
Messages arrive at the times of a point process. A message may be rejected
upon arrival or at any point during its sojourn in the system. If the message
eventually departs it is called succesful. Fix time $t$ and consider the time $D(t)$
of the last successful departure before $t$. Tag this message and let $A(t)$ be the
time it arrived in the system. Hence $A(t) \le D(t) \le t$, and $D(t)-A(t)$ is its
total sojourn time in the system. The quantity $\alpha(t)=t-A(t)$ is called
AoI at time $t$. As mentioned, the definition is quite general and doesn't
care about the particular system.
For further information on the topic see \cite{Yates12,Kosta17,KKZ19,KKZ21a,KKZ21b}
and references therein.

One would like to have certain deterministic functionals of the process $\alpha$ as
performance measures of the system. Typical examples are
 the expectation of the (possibly unique) stationary version of $\alpha$ at a fixed 
(and hence any) time, or the tail of its distribution. 
In \cite{KKZ19, KKZ21a, KKZ21b} we analyzed a variety of systems always aiming
at considering ones that have ``least'' AoI.
We complement these studies, in particular that of \cite{KKZ21a}, by examining the 
system of the following paragraph.

Consider a buffer of size $m$ where $m$ is a positive integer.
Position (cell) $1$ of the buffer is occupied by the message currently in service,
if any. Messages arrive at the times of a Poisson process with rate $\lambda$.
The service times of the messages are i.i.d.\ positive random
variables and independent of the arrival process.
A message receiving service is never interrupted by newly arrived ones.
Messages stored in cells $2,3, \ldots, m$ are in decreasing order of their
arrival times.
We use the letter $\PP_m$ for this system. 
For example, supposing that $m=3$,
and that messages labeled, say, $1,2,3,\ldots$,
arrive at times $T_1 < T_2 < T_3 <\cdots$, and if message $1$ has a very
long service time then here is what the contents of the  buffer look like,
while message $1$ is in the system:
$1 \varnothing \varnothing$,
$1 2 \varnothing$,
$1 3 2$,
$1 4 3$,
$1 5 4$, $\ldots$,
at times $T_1+\epsilon$, $T_2+\epsilon$, $T_3+\epsilon$, $T_4+\epsilon$, $T_5+\epsilon$, $\ldots$,
respectively, where $0<\epsilon\ll 1$.
When message $1$ departs all messages move one cell forward.

\section{The analysis}
We follow the notation of our previous papers.
The Poisson arrival process has points at times $T_k$, $k \in \Z$.
Message arriving at time $T_k$ has service time $\sigma_k$.
We denote the expectations of $T_2-T_1$ and $\sigma_1$ by $1/\lambda$ and
$1/\mu$, respectively.
We denote the 
the distribution of $\sigma_1$  by $G$.
We let $\hG$ be the Laplace transform of the probability measure $G$.
We assume that $\alpha(t)$, $t \in \R$, is stationary (which requires putting everything
on a suitable probability space).
Finally, we let $\tau$ be a random variable with distribution that of $\tau_k:=T_{k+1}
-T_k$, for any $k \in \Z \setminus \{0\}$.
We also ket  $\sigma$ be a random variable with distribution that of $\sigma_k$ for
any $k \in \Z$. Obviously, $\P(\tau > t) = e^{-\lambda t}$, $t \ge 0$,
and $\hG(s) = \E e^{-s \sigma}$, $s \ge 0$.

Let $S_n$, $n \in \Z$, be the sequence of succesful departures. 
We have that $S_n$ is a strictly increasing function of $n \in \Z$
and, for concreteness, we let $S_1$ be the smallest positive element of 
the sequence.
If we let $Q(t)$ be the number of messages in the buffer (the number
of occupied cells) at time $t$ and assume that $Q$ is right-continuous
(and stationary) then we set 
$K_n := Q(S_n)$. This is the number of occupied cells immediately after 
time $S_n$. Clearly, $K_n \in \{0,1,\ldots, m-1\}$ a.s., as the probability that
an arrival occurs exactly at the time of a successful departure is zero.
We observe that $\{K_n,S_n\}$, $n \in \Z$, is a Markov-renewal process.
In particular, $\alpha(0)$ is independent of 
$S_1-S_0$ given $K_0$.
The Markov chain $K_n$, $n \in \Z$, is stationary. We let
$\sP_{i,j} := \P(K_n=j|K_{n-1}=i)$, $i,j \in \{0,1,\ldots,m-1\}$,
and, for $m=3$, find that
\[
\sP =
\begin{pmatrix}
\P(\tau_1>\sigma) & \P(\tau_1\leq \sigma,\tau_1+\tau_2>\sigma) & \P(\tau_1+\tau_2\leq \sigma)\\
\P(\tau_1>\sigma) & \P(\tau_1 \leq \sigma,\tau_1+\tau_2>\sigma) & \P(\tau_1+\tau_2 \leq \sigma)\\
0 & \P(\tau_1>\sigma) & \P(\tau_1 \leq \sigma)
\end{pmatrix}
=
\begin{pmatrix}
\hG(\lambda) & -\lambda \hG'(\lambda) &  1-\hG(\lambda)+ \lambda \hG'(\lambda) \\
\hG(\lambda) & -\lambda \hG'(\lambda) &  1-\hG(\lambda)+ \lambda \hG'(\lambda) \\
0 & \hG(\lambda) & 1-\hG(\lambda) 
\end{pmatrix}.
\]
Here $\hG'$ is the derivative of $G$.
Let $\pi$ be the unique stationary distribution of $K$,
easily found from the equation $\pi = \pi \sP$:
\[
\pi_0 = 
\frac{\hG(\lambda)^2}{1+\lambda\hG'(\lambda)}, \quad
\pi_1 = \frac{\hG(\lambda)(1-\hG(\lambda))}{1+\lambda\hG'(\lambda)}, \quad
\pi_2 = \frac{1-\hG(\lambda)+\lambda\hG'(\lambda)}{1+\lambda\hG'(\lambda)}  .
\]

Noting that
\[
\alpha(t) = \alpha(S_0)+t-S_0, \quad S_0 \le t< S_1,
\]
and letting $\P^0$ be the Palm probability of $\P$ with respect to the
stationary point process $S_n$, $n \in \Z$, we use the Palm inversion formula  \cite{BB}
to obtain
\begin{align}
\E \alpha(0)   =
 \frac{ \E^0 \int_{S_0}^{S_1} \alpha(t) dt}{\E^0 (S_1-S_0)} 
 =\frac{ \E^0(\alpha(S_0)(S_1-S_0))
+\frac12 \E^0(S_1-S_0)^2}{\E^0 (S_1-S_0)}
= \frac{ \E^0 \alpha(0) S_1
+\frac12 \E^0 S_1^2}{\E^0 S_1},
\label{palm}
\end{align}
where we used that $\P^0(S_0=0)=1$. We figure out the terms
in this formula by conditioning on $K_0$:
\begin{equation}
\label{CID}
\E^0 (\alpha(0) S_1|K_0)  = 
\E^0[\alpha(0)| K_0] \, \E^0[S_1 |K_0].
\end{equation}

We explain how the terms in (\ref{palm}) are computed.
We focus on the  $m=3$ case.

\paragraph{Step 1.}
The following are elementary.
\begin{align*}
\begin{matrix*}[l]
\displaystyle \E(\sigma|\tau > \sigma)   = \frac{-\hG'(\lambda)}{\hG(\lambda)} 
& \displaystyle \qquad \E(\sigma|\tau\leq \sigma)  =\frac{\mu^{-1}+\hG'(\lambda)}{1-\hG(\lambda)} 
\\[5mm]
\displaystyle \E(\sigma|\tau_1+\tau_2\leq \sigma) = 
\frac{\mu^{-1}+\hG'(\lambda)-\lambda\hG''(\lambda)}
{1-\hG(\lambda) + \lambda \hG'(\lambda)} 
&\displaystyle \qquad 
\E(\sigma|\tau_1\leq \sigma,\tau_1+\tau_2>\sigma)  =
\frac{ \hG''(\lambda) }{ - \hG'(\lambda) }
\\[5mm]
\displaystyle \E(\tau_1|\tau_1+\tau_2\leq \sigma)  =  
\frac{ \lambda^{-1} - \lambda^{-1}\hG(\lambda)+\hG'(\lambda)
- \frac12 \lambda \hG''(\lambda)}
{1-\hG(\lambda) + \lambda \hG'(\lambda)} 
&\displaystyle \qquad 
\E (\tau_1|\tau_1\leq\sigma,\tau_1+\tau_2>\sigma)   = 
\frac{ \hG''(\lambda) }{ -2 \hG'(\lambda) }
\\[5mm]
\displaystyle  q :=\P(\tau_1+\tau_2>\sigma|\tau_1\leq \sigma)  = 
\frac{- \lambda \hG'(\lambda)
}{1-\hG(\lambda)} & 
\end{matrix*}
\end{align*}

\paragraph{Step 2.}
Using the fact that, under $\P^0$,
\[
S_1  \eqdist \tau {\bf 1}\{K_0=0\}+\sigma,
\]
we have
\[
\E^0 (S_1 | K_0) =  \lambda^{-1}  \1_{K_0=0}
+ \mu^{-1} ,
\]
and
\begin{align*}
\E^0 S_1 
& = \pi_0 \lambda^{-1}+\mu^{-1}
\\
\E^0 S_1^2 
& = \E \sigma^2 + \pi_0 2(\mu^{-1} + \lambda^{-1})\lambda^{-1}.
\end{align*}

\paragraph{Step 3.}
We condition further on (with hindsight) the information needed in
order to figure out $\alpha(0)$, namely, 
\[
\E^0 [\alpha(0)|K_0]
= \E^0 [\E^0(\alpha(0)| K_0, K_{-1}, K_{-2})|K_0],
\]
and then use the Markov property:
\begin{align}
\E^0 [\alpha(0)|K_0=\ell] = \sum_{i=0}^2\sum_{j=0}^2
\E^0 (\alpha(0)|K_{-2}=i,K_{-1}=j,K_{0}=\ell)  \,
\frac{\pi_i}{\pi_\ell} \sP_{\ell,j} \sP_{j,i}.
\label{alpha_K0}
\end{align}

The individual terms in \eqref{alpha_K0} must now be computed separately
by rolling up one's sleeves and treading carefully.

\paragraph{Step 4.}
We consider all possible values of $K_{-1}$ in the conditioning separately.

\paragraph{Step 4a.}
If $K_{-1}=0$ then $\alpha(0)$ is just 
the service time of the first message
arriving in $[S_{-1},S_0)$, irrespective of $K_{-2}$.
This translates into the explicit formula
\begin{equation}
\label{alpha_i0}
\E^0 (\alpha(0)|K_{-2}=i,K_{-1}=0,K_{0}=\ell)    = 
\begin{cases}
\E(\sigma|\tau>\sigma) ,  & \mbox{if $\ell=0$}\\
\E(\sigma|\tau_1\leq \sigma,\tau_1+\tau_2>\sigma) ,  & \mbox{if $\ell=1$}\\
\E(\sigma|\tau_1+\tau_2\leq \sigma) ,  & \mbox{if $\ell=2$}
\end{cases}.
\end{equation}

\paragraph{Step 4b.}
If $K_{-1}=1$ then things become a tad more complicated. We find that,
for $i\in\{0,1\}$,
\begin{equation}
\label{alpha_i1}
\E^0 (\alpha(0)|K_{-2}=i,K_{-1}=1,K_0=\ell) = 
\E(\tau_1|\tau_1\leq\sigma,\tau_1+\tau_2>\sigma) +  
\begin{cases}
\E(\sigma|\tau>\sigma) ,  & \mbox{if $\ell=0$}\\
\E(\sigma|\tau_1\leq \sigma,\tau_1+\tau_2>\sigma) ,  & \mbox{if $\ell=1$}\\
\E(\sigma|\tau_1+\tau_2\leq \sigma) ,  & \mbox{if $\ell=2$}
\end{cases}
,
\end{equation}
and, for $i=2$,
\begin{multline}
\label{alpha_21}
\E^0 (\alpha(0)|K_{-2}=2,K_{-1}=1,K_0=\ell) = 
q\E(\tau_1|\tau_1\leq \sigma,\tau_1+\tau_2 > \sigma) +  
(1-q)\E(\tau_1|\tau_1+\tau_2\leq \sigma) 
\\
 ~~~+ \E(\sigma|\tau>\sigma)+
\begin{cases}
\E(\sigma|\tau>\sigma) ,  & \mbox{if $\ell=0$}\\
\E(\sigma|\tau_1\leq \sigma,\tau_1+\tau_2>\sigma) ,  & \mbox{if $\ell=1$}\\
\E(\sigma|\tau_1+\tau_2\leq \sigma) ,  & \mbox{if $\ell=2$}
\end{cases}.
\end{multline}

\paragraph{Step 4c.}
If $K_{-1}=2$, we have, for $i \in \{0,1\}$,
\begin{equation}
\label{alpha_i2}
\E^0 (\alpha(0)|K_{-2}=i,K_{-1}=2,K_0=\ell) = 
\E(\tau_1|\tau_1+\tau_2\leq \sigma) +  
\begin{cases}
\E(\sigma|\tau>\sigma) ,  & \mbox{if $\ell=1$}\\
\E(\sigma|\tau\leq \sigma) ,  & \mbox{if $\ell=2$}
\end{cases}
,
\end{equation}
and, for $i=2$,
\begin{equation}
\label{alpha_22}
\E^0 (\alpha(0)|K_{-2}=2,K_{-1}=2,K_0=\ell)  = 
\E(\tau|\tau\leq \sigma) + 
\begin{cases}
\E(\sigma|\tau>\sigma) ,  & \mbox{if $\ell=1$}\\
\E(\sigma|\tau\leq \sigma) ,  & \mbox{if $\ell=2$}
\end{cases}.
\end{equation}

Of all the formulas in the last steps, only \eqref{alpha_i0} has been justified.
We now justify the remaining ones.

\begin{proof}[\bf\em Proof of \eqref{alpha_i1} and \eqref{alpha_21}]
$~$
\\
\begin{center}
\includegraphics[width=2.5in]{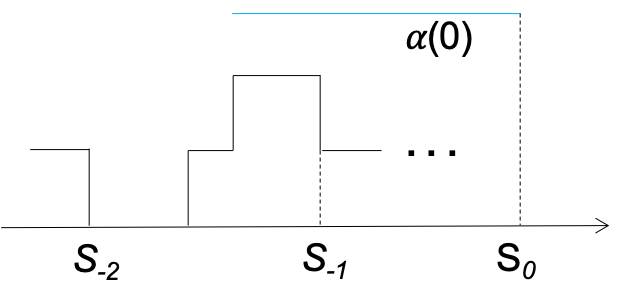} ~~~
\includegraphics[width=2.5in]{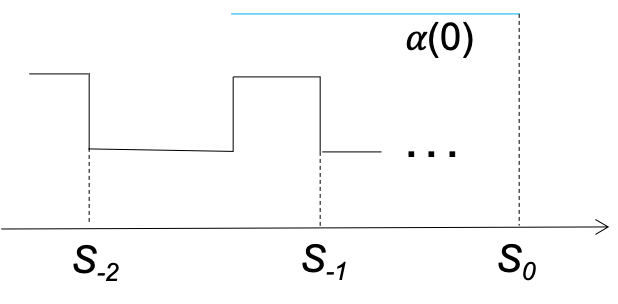} 
\captionof{figure}{Cases for $K_{-2}\in\{0,1\}$ when $K_{-1}=1$}\label{fig:alpha_1}
\end{center}
First see Figure \ref{fig:alpha_1}.
Given $K_{-1}=1$  we condition on 
$K_{-2}$. 
If $K_{-2}\in\{0,1\}$ then the message that departs at $S_0$ is the
only one that arrives in the service interval of
$[S_{-2},S_{-1})$. 
So  \eqref{alpha_i1} follows,
where the first term is by time-reversibility of the Poisson arrivals.

\begin{center}
\includegraphics[width=3in]{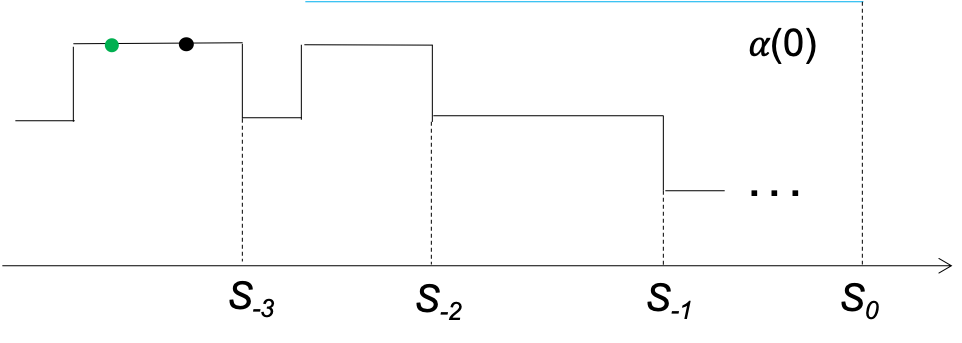} ~
\includegraphics[width=2.25in]{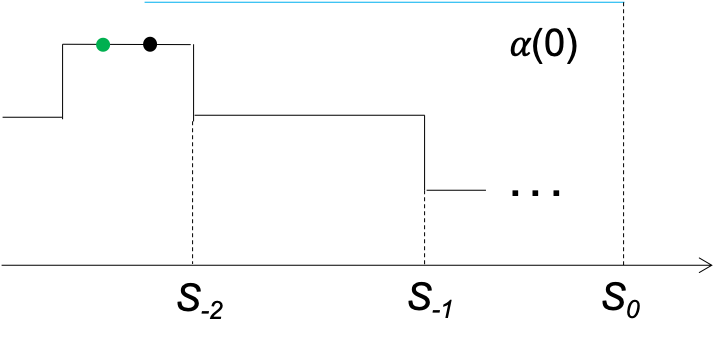}
\captionof{figure}{Cases for $K_{-2}=2$ when $K_{-1}=1$}\label{fig:alpha_1b}
\end{center}
The case given $K_{-2}=2,K_{-1}=1$ is interesting.
See Figure \ref{fig:alpha_1b} and
recall $q$ from Step 1.
$K_{-2}=2$ implies 
at least one
arrival in $[S_{-3},S_{-2})$, i.e.,
$\tau_1\leq \sigma$.
Generally, there is a geometric number $N\sim{\sf geom}(q)$
of consecutive intervals 
like $[S_{-3},S_{-2})$ of the left sub-figure 
with exactly one arrival.\footnote{That is,
$\P(N=k)=q^k(1-q)$ for $k=0,1,2,...$}
Just before these $N$ intervals
is an interval with more than one arrival, where
the second-last arrival (green dot) therein is serviced in $[S_{-1},S_0)$.
Recall that
the definition of AoI is the time since the arrival
of the most recently arrived message which has been completely served,
which at time $t=S_0=0$ 
is  {\em not} the (green) message served in $[S_{-1},S_0)$ in this case.
Note that $N=1$ in the left sub-figure,
$N=0$ in the right one, and $\P(N>0)=q$. 
Also, $S_{-1}-S_{-2} \eqdist (\sigma|\tau>\sigma)$.
So  \eqref{alpha_21} follows.
\end{proof}

\begin{remark}
Note that in the case of Figure \ref{fig:alpha_1b} at left,
the message being served in $[S_{-1},S_0)$ could be very stale -
a geometric number of fresher messages have been served before it.
This can also occur in $\PP_m$ for $m>3$ but not in $\PP_2$.
\end{remark}

\begin{proof}[\bf\em Proof of \eqref{alpha_i2} and \eqref{alpha_22}]
$~$
\\
\begin{center}
\includegraphics[width=2.14in]{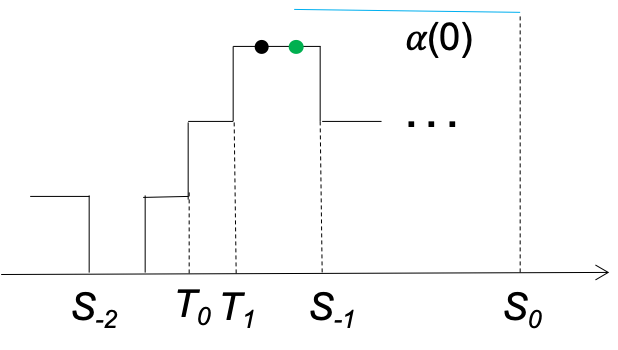} ~~~
\includegraphics[width=2.14in]{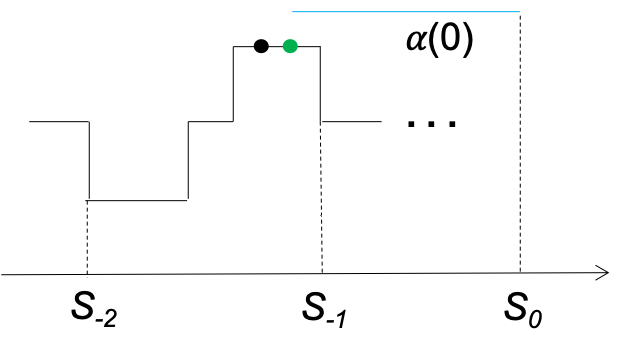} ~~~
\includegraphics[width=2.14in]{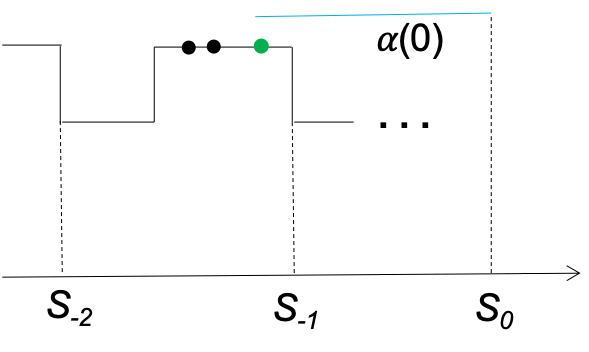}
\captionof{figure}{Cases for $K_{-2}\in\{0,1,2\}$  when $K_{-1}=2$}\label{fig:alpha_2}
\end{center}
Given $K_{-1}=2$, see Figure \ref{fig:alpha_2}.
If $K_{-2}=i$ for $i\in\{0,1\}$, there must be at least two message arrivals
during the service interval of $[S_{-2},S_{-1})$, i.e.,
at $T_0,T_1$. If there are only two arrivals, then
$\alpha(0)=S_0-T_1$, but there may be more than two
(black and green dots, where the green arrival is the one
served in $[S_{-1},S_0)$), in which case $\alpha(0)$ is as indicated
in the two left sub-figures. So \eqref{alpha_i2} holds,
where the first term is by time-reversibility of the
Poisson arrival process.
Similarly, if $K_{-2}=2,K_{-1}=2$  then there is at least one arrival
in $[S_{-2},S_{-1})$ and so  \eqref{alpha_22} holds.
\end{proof}

\section{The synthesis}
\label{synth}
Even for the case $m=3$, the analytical expression for 
$\E \alpha(0)$ is complicated and we shall not write it down. However,
we have all the ingredients (except patience) in order to 
tell the computer how to calculate it symbolically.
The focus is in formula \eqref{palm} which we repeat here:
\[
\tag{1}\label{palmagain}
\E \alpha(0)   
= \frac{ \E^0 \alpha(0) S_1
+\frac12 \E^0 S_1^2}{\E^0 S_1}.
\]
The terms on the right-hand side of \eqref{alpha_K0} are explicitly given by
the expressions \eqref{alpha_i0}--\eqref{alpha_22}
together with the elementary formulas of Step 1.
We thus know 
\[
\E^0 [\alpha(0)|K_0=\ell], \quad \ell=0,1,2.
\]
The quantities
\[
\E^0[ S_1|K_0=\ell], \quad \ell=0,1,2,
\]
are quite simple, see Step 2.
By the conditional independence formula \eqref{CID}, the quantity
\[
\E^0 \alpha(0) S_1 
= \sum_{\ell} \E^0(\alpha(0)|K_0=\ell)
\E^0 (S_1|K_0=\ell) \pi_\ell.
\]
is also known.
The first two moments of $S_1$ under $\P^0$ are also known 
from Step 2.
Automating the synthesis in the computer is not hard.
We only present a few plots below.

\section{Comparisons}

In the following, let $\alpha_m$ denote a random variable
distributed as $\alpha(0)$ for $\PP_m$.
We have
\[
\E\alpha_1 =(\lambda \hG(\lambda))^{-1},
\]
\begin{equation}  \label{P2_Aoi_mean}
	\mathbb{E}[\alpha_2] \;=\; \frac{1}{\mu}+ \frac{1}{\lambda}\left( 1-\hG(\lambda) + \lambda \hG'(\lambda) \right) + \frac{1}{\lambda}\, \frac{1}{\frac{\lambda}{\mu}+\hG(\lambda)} \left( \hG(\lambda) - \lambda \hG'(\lambda)
	+ \frac{1}{2} \lambda^2 \hG''(0) \right),
\end{equation}
from \cite{KKZ19,KKZ21a}.
For $\E \alpha_3$ we follow Section \ref{synth}.

For deterministic service times, 
i.e., $\hG(\lambda)= e^{-\lambda/\mu}$,
$\PP_2$ has been shown \cite{KKZ21a} to have lower mean AoI than $\PP_1$ (as well
as another system that we called  $\BB_1$ in \cite{KKZ19}). 
See Figure \ref{fig:det} which shows that $\PP_2$ has smaller
mean AoI than $\PP_1$ or $\PP_3$.

\begin{center}
\includegraphics[width=3.5in]{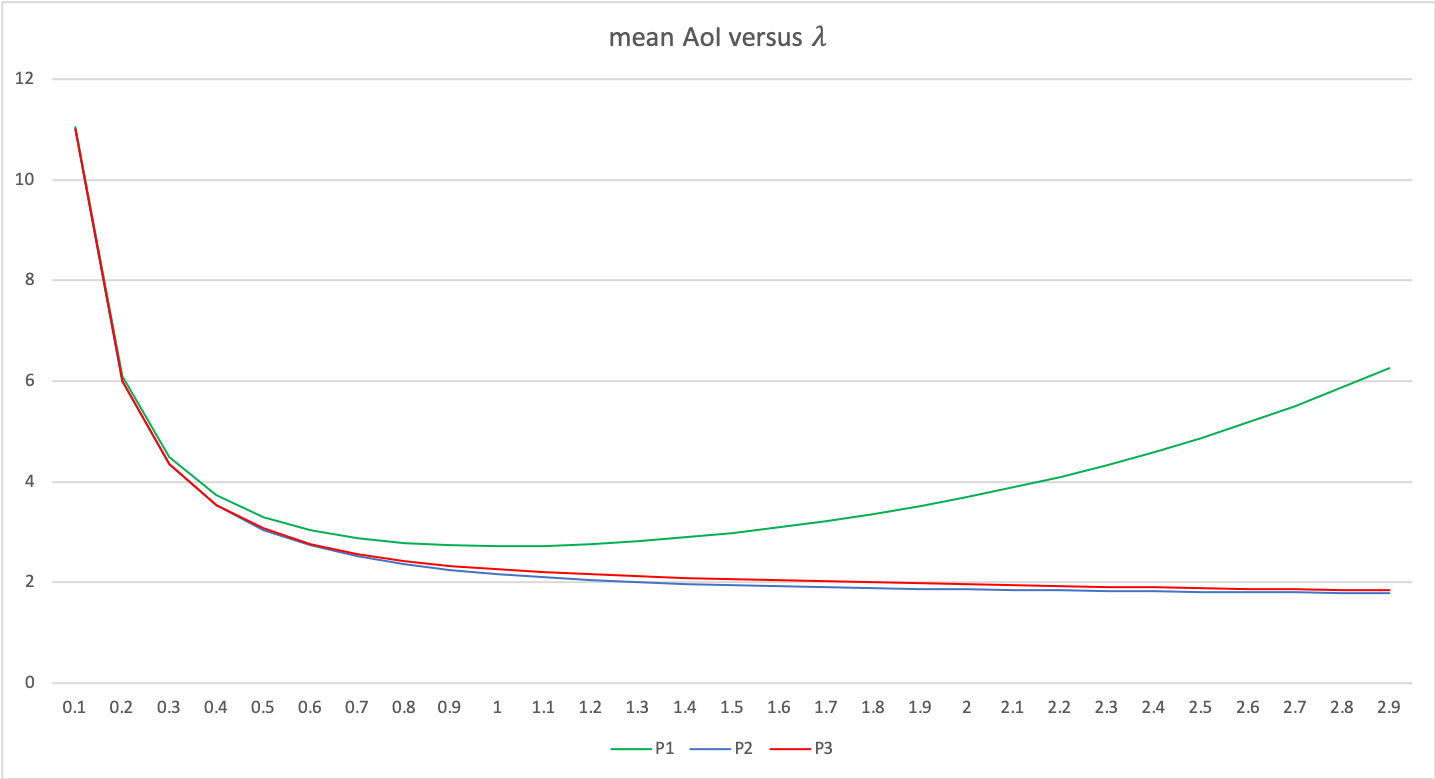}  
\captionof{figure}{For Poisson arrivals and deterministic
service times with $\mu=1$}\label{fig:det} 
\end{center}

For exponential service times, i.e.,
$\hG(\lambda)= \mu/(\lambda+\mu)$,
recall $\PP_1$ 
has lowest mean AoI \cite{Kosta17,KKZ21a},
consistent with Figure \ref{fig:exp}. Also,
$\PP_2$ has lower mean AoI than $\PP_3$.

\begin{center}
\includegraphics[width=3.5in,height=2in]{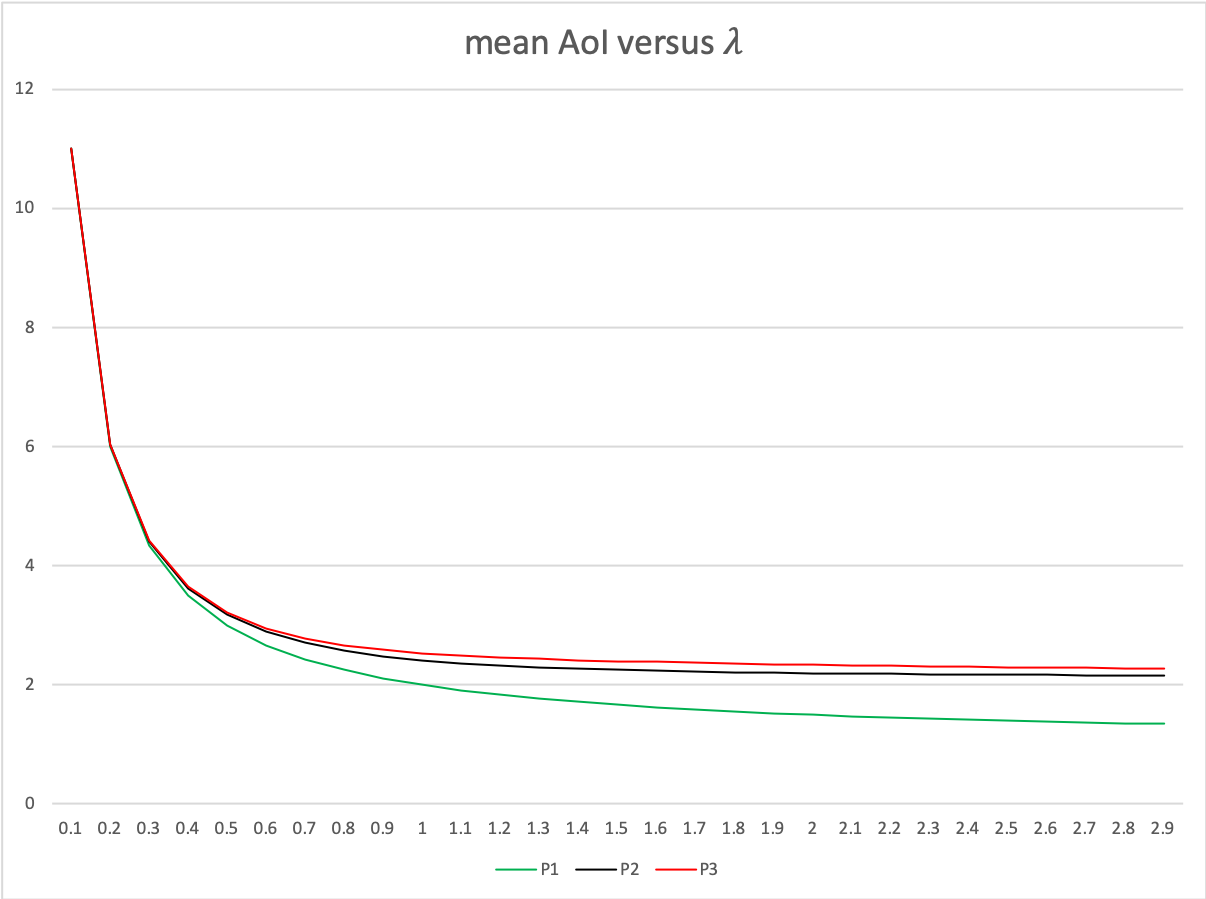}
\captionof{figure}{For Poisson arrivals and exponential
service times with $\mu=1$}\label{fig:exp} 
\end{center}

Note that $\PP_2$ and $\PP_3$ have the same
mean AoI in the limit as $\lambda\rightarrow\infty$.

\section{Afterword}

\no
1.\ The foregoing can be adapted in a straightforward
way for $\PP_m$ with $m>3$.
In particular, one observes that if the service times are deterministic then
$\E \alpha_2 < \E \alpha_3 < \E \alpha_4 < \cdots < \E \alpha_1$.
On the other hand, if the service times are exponential then
$\E \alpha_1< \E \alpha_2 < \E \alpha_3 < \E \alpha_4 < \cdots $.

\no
2.\ Even though we only showed how the expectation of $\alpha(t)$ can be 
computed, precisely the same algorithm can be followed if the 
Laplace transform (or even the distribution) of it is desirable.

\no
3.\ The case of renewal arrivals and exponential service
times can be  similarly analyzed. 

\bibliographystyle{plain}

\end{document}